\documentclass[aps,pra,nofootinbib,groupedaddress,twocolumn,final]{revtex4}

\usepackage{amssymb}
\usepackage{amsfonts}
\usepackage{amsmath}
\usepackage{verbatim}

\def\sb{{\sf b}}

\def\cH{{\cal H}}

\def\tr{{\rm tr}}

\newcommand\ket[1]{| #1 \rangle}

\newcommand\braket[2]{\langle #1|#2\rangle}
\newcommand\ketbra[2]{|#1\rangle\langle #2|}

\newcommand\trnorm[1]{\| #1 \|_1}

\begin{document}

\title{How to Build Unconditionally Secure Quantum Bit Commitment
  Protocols\footnote{NOTE: This paper and the
  previous two (quant-ph/0305142 and quant-ph/0305143) together provide a detailed description of various
  gaps in the QBC ``impossibility proof,'' as well as security
  proofs for four different protocols, QBC1 QBC2, QBC4, and QBC5. They also explain and correct some of the claims on my
  previous QBC protocols. Among various minor improvements over v2,
  the main change in this v3 involves the deletion of QBC2.1p and an
  expanded discussion of Type 3 protocols.}}

\author{Horace P. Yuen}\email{yuen@ece.northwestern.edu}
\affiliation{Department of Electrical and Computer Engineering, Department of
Physics and Astronomy, Northwestern University, Evanston, IL
60208-3118, USA}




\begin{abstract}
The ``impossibility proof'' on unconditionally secure quantum bit
commitment is critically analyzed. Many possibilities for obtaining a
secure bit commitment protocol are indicated, purely on the basis of
two-way quantum communications, which are not covered by the
impossibility proof formulation. They are classified under six new
types of protocols, with security proofs for specific examples on four
types. Reasons for some previously failed attempts at obtaining secure
protocols are also indicated.
\end{abstract}

\maketitle




\section{Introduction}\label{sec:intro}

Bit commitment is a kind of a cryptographic protocol that can serve as
a building block to achieve various cryptographic objectives, such as
user authentication. There is a nearly universal acceptance of the
general impossibility of secure quantum bit commitment (QBC), taken to
be a consequence of the Einstein-Podolsky-Rosen (EPR) type
entanglement cheating which supposedly rules out QBC and other quantum
protocols that have been proposed for various cryptographic objectives
\cite{bc}.  In a {\it bit commitment} scheme, one party, Adam,
provides another party, Babe, with a piece of evidence that he has
chosen a bit \sb\ (0 or 1) which is committed to her.  Later, Adam
would {\it open} the commitment by revealing the bit \sb\ to Babe and
convincing her that it is indeed the committed bit with the evidence
in her possession and whatever further evidence Adam then provides,
which she can {\it verify}.  The usual concrete example is for Adam to
write down the bit on a piece of paper, which is then locked in a safe
to be given to Babe, while keeping for himself the safe key that can
be presented later to open the commitment.  The scheme should be {\it
binding}, i.e., after Babe receives her evidence corresponding to a given bit
value, Adam should not be able to open a different one and convince
Babe to accept it. It should also be {\it concealing}, i.e., Babe
should not be able to tell from her evidence what the bit \sb\ is.
Otherwise, either Adam or Babe would be able to cheat successfully.

In standard cryptography, secure bit commitment is to be achieved
either through a trusted third party, or by invoking an unproved
assumption concerning the complexity of certain computational
problems.  By utilizing quantum effects, specifically the intrinsic
uncertainty of a quantum state, various QBC schemes not
involving a third party have been proposed to be unconditionally
secure, in the sense that neither Adam nor Babe could cheat with any
significant probability of success as a matter of physical laws.  In
1995-1996, a supposedly general proof of the impossibility of
unconditionally secure QBC, and the insecurity of previously proposed
protocols, were presented \cite{may1}-\cite{lc3}.  Henceforth it has
been generally accepted that secure QBC and related objectives are
impossible as a matter of principle \cite{lo}-\cite{sr}.

There is basically just one impossibility proof (IP), which gives the EPR
attacks for the cases of equal and unequal density operators that Babe
has for the two different bit values.  The proof purports to show that
if Babe's successful cheating probability $P^B_c$ is close to the
value 1/2, which is obtainable from pure guessing of the bit value,
then Adam's successful cheating probability $P^A_c$ is close to the
perfect value 1.  This result is stronger than the mere impossibility
of unconditional security, namely that it is impossible to have both
$P^B_c \sim 1/2$ and $P^A_c \sim 0$.  The impossibility proof
describes the EPR attack on a specific type of protocols, and then
argues that all possible QBC protocols are of this type.

Typically, one would expect that a proof of impossibility of carrying
out some thing X would show that any possible way of doing X would
entail a feature that is logically contradictory to given principles,
as, for example, in the cases of quantum no-cloning \cite{wz,yue1} and
von Neumann's no-hidden-variable theorem \cite{neu}.  In the present
case, one may expect a proof which shows, e.g., that {\it any} QBC
protocol that is concealing is necessarily not binding. It is
important for this purpose that the QBC protocol
formulation be all-inclusive.  In the
absence of a proof that all possible QBC protocols have been included
in its formulation, any impossibility proof is at best
incomplete. Thus, {\it a priori}, there can be {\it no} general
impossibility proof without a mathematical characterization or a
definition of all QBC protocols. Within the framework of two-way
quantum communications between Adam and Babe with no further
constraints, which is the setting of the IP and its EPR attack, no
such definition has ever been presented. Although one can judge
whether or not a specific protocol is a QBC protocol, similar to whether a
specific process of computation is ``algorithmic'' or not, it
appears prohibitively difficult to characterize mathematically all QBC
protocols. Just as there is no Church-Turing theorem, just a
Church-Turing thesis, there can be no impossibility theorem without a
mathematical definition of a QBC protocol.

More concretely, there are many typces of QBC protocols that are not
captured by the IP formulation and differ from each other, similar to the
existence of several types of algorithmic processes different from a
Turing machine. Some of these types were described by this author
previously \cite{yuea}-\cite{yuec}, and secure protocols can actually be
found among them. As those papers, and this one, should make clear,
even if secure QBC were impossible in all of these types, the
impossibility proof for each of them would be different and would bear
no resemblance to the well-known IP. These different types of
protocols arise because only certain techniques of protocol design,
such as certain use of classical randomness in a quantum protocol, are
included in the IP formulation, which does not show that all possible
techniques have been included. Even just for classical randomness, the
different ways it could affect a QBC protocol are not properly
accounted for. In this paper, a systematic description of some gaps in
the IP, and the corresponding opportunity for six new protocol types,
would be identified and elaborated upon. The original IP formulation
would be named Type 0, with the new ones named Type 1 to Type 6.

In Section \ref{sec:impproof}, the impossibility proof is reviewed and
its scope delimited. The basis of previous incorrect claims on the
security of several different QBC protocols would be discussed. See
also Appendix A. In Section \ref{sec:ipprobs}, the incompleteness of
the IP is analyzed from a variety of angles that lead to at least six
different types of protocols not covered by the IP formulation. They
are discussed in Section \ref{sec:types} with security proofs
given for QBC1 and QBC2. The security proofs for QBC4 and QBC5 are
given in Refs.~\cite{yued} and \cite{yuee}.

\section{The impossibility proof:  Type 0 protocols}\label{sec:impproof}

The impossibility proof, in its claimed generality, has never been
systematically spelled out in one place, but the essential ideas that
constitute this proof are generally agreed upon
\cite{may2}-\cite{sr}.  The formulation and the proof can be cast as
follows.  Adam and Babe have available to them two-way quantum
communications that terminate in a finite number of exchanges, during
which either party can perform any operation allowed by the laws of
quantum physics, all processes ideally accomplished with no
imperfection of any kind.  During these exchanges, Adam would have
committed a bit with associated evidence to Babe.  It is argued that,
at the end of the commitment phase, there is an entangled pure state
$\ket{\Phi_\sb}$, $\sb \in \{0,1\}$, shared between Adam who
possesses state space $\cH^A$, and Babe who possesses $\cH^B$.  For
example, if Adam sends Babe one of $M$ possible states $\{
\ket{\phi_{\sb i}} \}$ for bit \sb\ with probability $p_{\sb i}$, then
\begin{equation}
\ket{\Phi_{\sb }} = \sum_i \sqrt{p_{\sb i}}\ket{e_i}\ket{\phi_{\sb i}}
\label{eq:entstate}
\end{equation}
with orthonormal $\ket{e_i} \in \cH^A$ and known $\ket{\phi_{\sb i}}
\in \cH^B$.  Adam would open by making a measurement on $\cH^A$, say
$\{ \ket{e_i} \}$, communicating to Babe his result $i_0$ and $\sb$;
then Babe would verify by measuring
the corresponding projector $\ketbra{\phi_{\sb i_0}}{\phi_{\sb i_0}}$ on $\cH^B$,
accepting as correct only the result 1. More generally, one may
consider the whole $\ket{\Phi_\sb}$ of (\ref{eq:entstate}) as the
state corresponding to the bit $\sb$, with Adam sending $\cH^A$ to
Babe upon opening, so she can verify by projection measurement on $\ketbra{\Phi_\sb}{\Phi_\sb}$.

When classical random numbers known only to one party
are used in the commitment, they are to be replaced by corresponding
quantum entanglement purification.  The commitment of $\ket{\phi_{\sb
    i}}$ with probability $p_{\sb i}$ in (\ref{eq:entstate}) is, in
fact, an example of such purification. An example involving Babe may
be a protocol \cite{yue2}-\cite{yue4} where $\ket{\phi_{\sb i}}$ in
(\ref{eq:entstate}) is to be obtained by Adam applying unitary
operations $U_{\sb i}$ on state $\ket{\psi_k} \in \cH^{B_1}$ sent to him
by Babe with probability $\lambda_k$, $k \in K$, where $|K| < \infty$.
Generally, for any random $k$ used by Babe, it is argued that from the
doctrine of the ``Church of the Larger Hilbert Space'' \cite{gl}, it is to be
replaced by the purification $\ket{\Psi}$ in $\cH^B_\alpha \otimes \cH^B_\beta$,
\begin{equation}
\ket{\Psi} = \sum_k \sqrt{\lambda_k} \ket{\psi_k}\ket{f_k},
\label{eq:purif}
\end{equation}
where $\ket{\psi_k} \in \cH^B_\alpha$ and the $\ket{f_k}'s$ are complete orthonormal in $\cH^B_\beta$ kept
by Babe while $\cH^B_\alpha$ would be sent to Adam.  With such purification, it is claimed that any protocol
involving classical secret parameters would become quantum-mechanically
determinate, i.e., the shared state $\ket{\Phi_\sb}$ at the end of
commitment is completely known to both parties.  Note that, from
(\ref{eq:purif}), this means that both $\{\lambda_k\}$ and
$\{\ket{f_k}\}$ are taken to be known exactly to both Babe {\it and
Adam}. The possibility that one can always purify a classically random
situation as in (\ref{eq:purif}) has never been proved, especially how
it may be combined with the following (\ref{eq:unitop}). It is elaborated later in this paper and in Ref.~\cite{yuee}, in connection
with QBC1 and QBC5, that this is generally not possible. 

Why should Adam and Babe share a pure state at the end of commitment?
Any measurement followed by a unitary operation $U_l$ depending on the
measurement result $l$ can be equivalently described by an overall
unitary operator.  Thus, if the orthonormal $\{ \ket{g_l} \}$ on
$\cH^{C_2}$ is measured with result $l$, and then $U_l$ operates on $\cH^{C_1}$, it is
equivalent to the unitary operation
\begin{equation}
U = \sum_l U_l \otimes \ketbra{g_l}{g_l}
\label{eq:unitop}
\end{equation}
on $\cH^{C_1} \otimes \cH^{C_2}$.  It is claimed that any actual
measurement during commitment can be postponed until the opening and the
verification phases of the protocol without affecting the protocol in
any essential way. Actually, if the measurement result is announced during
commitment there is no need for (\ref{eq:unitop}) because the protocol
state is just indexed by the measurement result $\ell$ known to both
parties. (With the delayed measurement description, the
cheating and the opening would be quite involved and hard to describe
when the possibility of aborting a protocol is allowed. They have
never been explicitly spelled out.)  In order to maintain quantum
determinacy, the exact $\{ \ket{g_l} \}$ in
(\ref{eq:unitop}) are taken to be known to both parties. Let us use
$k$ to denote Babe's secret parameter, and $i$ to denote Adam's secret
parameter, such as the $i$ with probabilities $\{p_i\}$ in
(\ref{eq:entstate}). These
crucial {\it assumptions} of openly known $\{p_i\}$, $\{ \lambda_k \}$, $\{
\ket{f_k} \}$, and $\{ \ket{g_l} \}$ are made in the
impossibility proof through the use of known fixed quantum computers
or quantum machines for data storage and processing by either party
\cite{may2}, \cite{lc3}, \cite[Appendix]{bcml2}, even though the control of such machines belongs
only to one of the parties. As it turns out, to cheat successfully, Adam does not
need to know $\{ \ket{g_l}\}$ in a careful formulation and he does not
need to know $\{ \ket{f_k}\}$ in one general class of protocols
\cite{yue4}. However, he does need to know $\{ \ket{f_k}\}$ in
general, a fact which is exploited in our Type 4 protocols. The
general possibility of such quantities being unknown or classically
random to Adam is exploited in many of our protocols. A general
limitation on the quantum purification of classical randomness is
described in our discussion of Type 3 protocols in Section \ref{ssec:type3}.

Protocols of the form (\ref{eq:entstate}), where $\{ \ket{\phi_{\sb
    i}}\}$ are just sent from Adam to Babe, will be called {\it
    single-stage} protocols. In a {\it multiple-stage} protocol,
    $\ket{\phi_{\sb i}}$ becomes, for $i
    = \{i_1, \ldots, i_n\}$ and $k = \{ k_1,\ldots,k_{n-1}\}$ with $2n-1$ stages in total \cite{sr}
\begin{equation}
\ket{\phi_{\sb ik}} = U^A_{\sb i_n} \ldots U^A_{\sb i_2} U^B_{k_1}
U^A_{\sb i_1} \ket{\phi_0}.
\label{eq:multistage}
\end{equation}
The initial state $\ket{\phi_0}$ is openly known, and the alternate
possible unitary operations by Adam and Babe, together with their
respective probabilities, are also taken to be openly known. If the
action is initiated by Babe instead of Adam, (\ref{eq:multistage}) can
be replaced by, for a $2n$-stage protocol,
\begin{equation}
\ket{\phi_{\sb ik}} = U^A_{\sb i_n} \ldots U^B_{k_1}\ket{\phi_0}.
\label{eq:multistage_babe}
\end{equation}
Purification of the random $U$'s is to be carried out as in
(\ref{eq:purif}). Thus, a multi-stage protocol is equivalent to one of
the form
\begin{equation}
\ket{\Phi_{\sb}} = \sum_{ik}\sqrt{p_{\sb i}\lambda_k} \ket{e_i}
\ket{f_k} \ket{\phi_{\sb ik}},
\label{eq:multistage_equiv}
\end{equation}
where $\{ p_{\sb i}\}$, $\{ \lambda_k \}$, $\{ \ket{\phi_{\sb ik}}\}$
are openly known, $\ket{e_i} \in \cH^A$ controlled by Adam, $\ket{f_k}
\in \cH^{B_1}$ controlled by Babe, and $\ket{\phi_{\sb ik}} \in
\cH^{B_2}$ is the evidence Babe possesses at the end of
commitment. As in the case of quantum coin-tossing \cite{dk}
formulation, in the IP the whole state space $\cH^{B_2}$ is supposed
to be passed on during each stage. As described later, this {\em misses} the
nonuniqueness associated with passing back a portion of the space
during commitment or opening, as in QBC1, and the problem of the very
possibility of purification (\ref{eq:purif}), as in QBC5. By writing $\ket{\phi_{\sb i}} = \sum_k
\sqrt{\lambda_k} \ket{f_k}\ket{\phi_{\sb ik}}$,
(\ref{eq:multistage_equiv}) is also of the form (\ref{eq:entstate})
with $\cH^B = \cH^{B_1} \otimes \cH^{B_2}$, and a multi-stage protocol
is claimed to be equivalent to a single-stage one. Indeed, it is
alternatively argued that $\ket{\Phi_\sb}$ is always openly known at
the end of commitment in any multi-stage protocol with the use of
purification. Thus, it can always be represented by
(\ref{eq:entstate}) with all the quantities involved openly known.

With such a formulation, Babe can try to identify the bit from $\rho^B_\sb$, the
marginal state of $\ket{\Phi_\sb}$ on $\cH^B$, by performing an
optimal quantum measurement that yields the optimal cheating
probability $\bar{P}^B_c$ for her. Adam cheats by committing
$\ket{\Phi_0}$ and making a measurement on $\cH^A$ to open $i_0$ and
$\sb=1$.  His probability of successful cheating is computed through
$\ket{\Phi_\sb}$, his particular measurement, and Babe's verifying
measurement; the one optimized over all of his possible actions will
be denoted $\bar{P}^A_c$.  For a fixed measurement basis, Adam's
cheating can be described by a unitary operator $U^A$ on $\cH^A$. His
general EPR attack goes as follows.  If the protocol is
perfectly concealing, i.e, $\bar{P}^B_c = 1/2$, then $\rho^B_0 = \rho^B_1$.  By writing
$\ket{\Phi_\sb}$ as the Schmidt decomposition on $\cH^A \otimes
\cH^B$,
\begin{equation}
\ket{\Phi_\sb} = \sum_j \sqrt{\tilde{p}_j} \ket{\tilde{e}_{\sb j}}
\ket{\tilde{\phi}_j},
\label{eq:schmidt}
\end{equation}
where $\ket{\tilde{\phi}_j}$ are the eigenvectors of $\rho^B_\sb$ and
$\{ \ket{\tilde{e}_{\sb j}}\}$ for each $\sb$ are complete orthonormal in
$\cH^A$, it follows that Adam can obtain $\ket{\Phi_1}$ from
$\ket{\Phi_0}$ by a local cheating transformation $U^A$ that brings
$\{ \ket{e_{0j}} \}$ to $\{ \ket{e_{1j}}\}$. Thus his optimum cheating
probability is $\bar{P}^A_c =1$ in this case. More generally, when
Babe checks $\ket{\Phi_\sb}$ on $\cH^A \otimes \cH^B$, Adam still just
cheats by applying a local transformation $U^A$ to turn $\ket{\Phi_0}$
to $\ket{\Phi_1}$, although the terminology of EPR attack then becomes
somewhat misleading.

For unconditional, rather than perfect, security, one demands that
both cheating probabilities $\bar{P}^B_c - 1/2$ and $\bar{P}^A_c$ can
be made arbitarily small when a security parameter $n$ is increased
\cite{may2}.  Thus, {\it unconditional security} is quantitatively expressed
as
\begin{equation}
({\rm US}) \qquad \lim_n \bar{P}^B_c = \frac{1}{2},\quad \lim_n
  \bar{P}^A_c = 0.
\label{eq:us}
\end{equation}
The condition (\ref{eq:us}) says that, for any $\epsilon > 0$, there
exists an $n_0$ such that for all $n > n_0$, $\bar{P}^B_c - 1/2 <
\epsilon$ and $\bar{P}^A_c < \epsilon$, to which we may refer as
$\epsilon$-{\it concealing} and $\epsilon$-{\it binding}.  These
cheating probabilities are to be computed purely on the basis of
logical and physical laws, and thus would survive any change in
technology, including an increase in computational power.  In general,
one can write down explicitly
\begin{equation}
\bar{P}^B_c = \frac{1}{4}\left(2 + \trnorm{\rho^B_0 -
  \rho^B_1}\right),
\label{eq:barpc}
\end{equation}
where $\trnorm{\cdot}$ is the trace norm, $\trnorm{\tau} \equiv \tr
(\tau^\dag \tau)^{1/2}$ for a trace-class operator $\tau$, but the
corresponding $\bar{P}^A_c$ is more involved. However, it may be shown
that it satisfies \cite{yueb}
\begin{equation}
4(1-\bar{P}^B_c)^2 \le \bar{P}^A_c \le 2 \sqrt{\bar{P}^B_c
  (1-\bar{P}^B_c)}.
\label{eq:cheatbounds}
\end{equation}
The lower bound in (\ref{eq:cheatbounds}) yields the following
impossiblity result given by the IP, 
\begin{equation}
\lim_n \bar{P}^B_c = \frac{1}{2} \,\, \Rightarrow
  \,\, \lim_n \bar{P}^A_c = 1
\label{eq:ip}
\end{equation}
within its formulation \cite{may1,yue2}. Condition (\ref{eq:us}) or
(\ref{eq:ip}) is a continuity statement different from a point
statement $\bar{P}^B_c = 1/2 \Rightarrow \bar{P}^A_c =1 $. Note that the impossibility
proof makes a stronger statement than the mere
impossibility of (US), i.e., (\ref{eq:ip}) is stronger than
(\ref{eq:us}) not being possible.

There have been quite a few incorrect claims on obtaining US QBC
protocols, both before and after the appearance of the IP. In
particular, two of the various approaches that were pursued by the
present author do not work for reasons associated with the IP (see
Appendix A for a summary). In the
first case, also proposed in different forms by several others, simple
use of classical randomness by Babe supposedly leads to different
cheating transformations for Adam dependent on such randomness, hence
a binding protocol is obtained after averaging over such randomness
that has to be carried out in evaluating $\bar{P}^A_c$. The
purification (\ref{eq:purif}) is not attended to, in view of the
``equivalence'' between classical and quantum randomness via the
``Church of the Larger Hilbert Space'' doctrine. However, this
doctrine, often used in the IP as in (\ref{eq:purif}), is {\em
incorrect}. One simple way to see that is to observe if Adam does not
entangle the classical randomness, e.g., if Adam sends $\ket{\phi_{\sb
i}}$ with probability $p_{\sb i}$ instead of (\ref{eq:entstate}), he
cannot launch entanglement cheating though $\rho^B_0 = \rho^B_1$ still
applies. Even when the entanglement purification has been carried
out, (\ref{eq:purif}) is equivalent to classically random $\{
\ket{\psi_k}\}$ only if the measurement of $\{ \ket{f_k} \}$ is first
performed on $\cH^{B_2}$. Otherwise, the off-diagonal elements in
$\rho^B_\sb$ involving $\ketbra{f_k}{f_{k'}}$ may lead to better
$\bar{P}^B_c$ in Babe's cheating measurement on $\cH^{B_1} \otimes
\cH^{B_2}$, as compared to the case of purely classically random $\{
\ket{\psi_k} \}$ with zero off-diagonal elements. As a specific
example, consider the protocol preceding QBC5p discussed in the
beginning of Ref.~\cite{yuee}. It is perfectly concealing if Babe does
not entangle, but not if she does. Other examples not involving
teleportation can also be given. Thus, there is {\em no} equivalence
between classical randomness and quantum purification. It is the
possibility of entanglement cheating by Babe, not Church of the Larger
Hilbert Space, which dictates that (\ref{eq:entstate})-(\ref{eq:purif})
is the correct representation in siuch a situation. Under such a
stronger concealing condition, compared to just classically random $\{
\ket{\psi_k}\}$, Adam may indeed cheat in accordance with IP,
depending on the protocol.

The second failed approach involves various attempts to turn a pure
$\ket{\Phi_\sb}$ into a mixed one through Adam's action during the
commitment phase before opening. The IP argues that a pure
$\ket{\Phi_\sb}$ can always be maintained in principle via perfect quantum
computation. While there is no mathematical formalization
on this issue that may lead to a rigorous proof, my different
attempts indeed lead to different countermeasures by Adam, and I do
not see what next attempt to try in this approach that may appear to
have a possibility of success. However, this strategy works on Babe's
entanglement and leads to our QBC1. The reason is directly connected
to the point of the last paragraph, namely that Babe's measurement on
$\{ \ket{f_k} \}$ first and then on $\cH^{B_2}$ can lead to a
concealing protocol even though there may be a measurement on $\cH^{B_1} \otimes
\cH^{B_2}$ with which Babe can cheat. Thus, Babe's entanglement may be
effectively ``destroyed'' through, e.g., Adam's questioning during
commitment. See the following discussions related to QBC1 in Sections
\ref{sec:ipprobs} and \ref{sec:types} for details.

\section{Problems of the impossibility proof}\label{sec:ipprobs}

A plausible first reaction to the impossibility proof is: why are all
possible QBC protocols covered by its formulation? More precisely, how
may one capture mathematically the necessary feature of an
unconditionally secure QBC protocol in a precise definition that is
required for the formulation and proof of a mathematical theorem that
says such a protocol is impossible? No such definition is
available. An analogy to a QBC protocol is an ``effectively computable''
function, a function whose value for any specific argument can be
``mechanically'' obtained in a finite number of steps without the
intervention of ``intelligence.''  The well-known Church-Turing thesis
says that any effectively computable function can be computed
recursively or by a Turing machine. It can be cast as an impossibility
statement: there is no effective procedure that cannot be simulated by
a Turing machine. It was found that a function that
can be computed by a method that is clearly effective, such as Post
machines or Markov algorithms, is indeed also Turing-computable.
However, nobody calls the Church-Turing thesis the Church-Turing
theorem.  This is because there is no mathematical definition of an
effective procedure.  The logical possibility is open that someday a
procedure is found that is intuitively or even physically effective,
but which can compute a nonrecursive arithmetical function.

Thus, in the absence of a precise definition of a QBC protocol, one
would have at best an ``impossibility thesis,'' not an impossibility
{\em theorem}.  (This view was emphasized to the author by Masanao
Ozawa.) It is often difficult, if not impossible, to capture
precisely by means of a mathematical characterization a given kind of physical operations that have unambiguous
common-sense meaning. For example, there is no
definition that would characterize all classical cryptographic
protocols, say for bit commitment.  It is at least not clear why a definition in the more general
quantum case can ever be found. Just as
there appear to be many different forms of effective procedures, there are
many different QBC protocol types that appear not to be captured by
the IP formulation.  To uphold just the
``impossibility thesis,'' one would need to prove that US QBC is
impossible in each of these types.

The problem of characterizing mathematically all QBC protocols,
although quite difficult, does not seem to be as hopeless as that for an
effectively computable function, if one believes that ``bit commitment
protocol'' is less ambiguous than ``effective procedure,'' even
though both concepts can presumably be recognized to be applicable or not when a
particular instance is presented. In particular, the
framework of two-way quantum communication, or the Yao model
\cite{yao}, without any constraints of relativity or superselection
rules but with the possibility of the protocol being aborted as a result of cheating
detection before opening, is an appropriate general setting. (The Yao
model allows actual measurements during rounds but is often
interpreted \cite{lc3} to imply
(\ref{eq:multistage_babe})-(\ref{eq:multistage_equiv}) used in
coin-tossing formulation. That would exclude the possibility of
sending back only part of a product space, which is utilized in many of
our US protocols.)  It is
sometimes argued that every proof has to presuppose a ``model,'' but
the question is whether the model used in the IP is general enough to
capture all clear-cut QBC protocols within the above framework. It is also
sometimes argued that the ``community of experts in the field''
have already agreed on a ``definition'' of what constitutes a QBC
protocol, which would rule out some of our Type 3 protocols. But the question
is why a clear-cut QBC protocol should be ruled out by
legislation. Note that ``definition'' in this context does not mean an
arbitrary choice of terminology, but a mathematical characterization
of all instances where the concept is applicable. Observe also that
this characterization problem does not arise in security proofs,
because one should be able to exhaust all possible types of attack
given a specific protocol.

The most important instances of incompleteness of the IP and quantum
coin-tossing formulations, as presently
understood
by the author, are listed under four categories in the following. Some
of these have been discussed previously \cite{yuea}-\cite{yuec},
\cite{yue2}. Some new protocol types made possible by such gaps are
discussed in Section \ref{sec:types}.\\

\paragraph{Freedom of Operation ---} In a two-party situation where
either one can do anything, constrained only by physical laws, and has
only his/her own interest to protect, neither can be trusted to be
honest if an operation would lead to his/her advantage without
penalty. On the other hand, a party is supposed to strive to achieve the aim of the
protocol if his/her own security against cheating by the other party
can be assured. These obvious considerations are codified as the
Libertarian Principle and the Intent Principle of protocol formation,
further elaborated in Ref.~\cite{yueb}. The resulting freedom of action
by either party is not accounted for in the QBC IP formulation, nor in
the mathematical formulations of quantum coin-tossing protocols
\cite{dk}.

\begin{itemize}
\item {\it Honesty and Cheating ---} In a multi-stage
  protocol, where a state space is passed between Adam and Babe in
  rounds for operations, as
  (\ref{eq:multistage_babe})-(\ref{eq:multistage_equiv}), either party can
  substitute an entirely different space of his/her own at any
  stage. The possible advantage is clear in coin-tossing, and examples
  were given in Ref.~\cite{yue4} on bit commitment. There is no mechanism
  built into the protocol formulation to prevent such
  cheating. Fortunately, this problem can be alleviated in two
  different ways \cite{yuee} including cheating detection and the
  possibility of aborting the protocol, with perhaps a penalty imposed
  on the party that got caught cheating with the use
  of an ensemble, as described in Ref.~\cite{yuee} for QBC5. However, there is still a lot of
  freedom left which has not been accounted for, some to be discussed
  in the following points (c) and (d).
\item {\it Random vs.~Nonrandom Secret Parameters ---} Suppose a
  protocol has the property that it is concealing for every possible
  legal operation by Babe that can be checked as mentioned above. Then
  Babe is free to choose any such operation (or state) with whatever
  probability distribution unknown to Adam. This freedom, codified as
  the Secrecy Principle \cite{yueb}, \cite{yuec}, is a simple corollary of the
  Libertarian Principle and the Intent Principle. It directly
  {\it contradicts} the IP claim that a state is always openly known at the
  end of commitment in a QBC protocol. One consequence of this freedom
  is that Adam's cheating transformations may depend on exactly what
  Babe's choice is in order to succeed. Indeed, he may not even have a
  single density operator representation for each $\sb$ due to the
  difference between a random parameter and an unknown parameter, a
  distinction well-known in statistics. Our Type 3 and Type 6
  protocols arise from this freedom, but no concrete protocol has been
  found in these types that can be proved unconditionally secure. On
  the other hand, with additional features one can construct secure protocols
  utilizing this freedom, as in our QBC1, QBC2, and QBC4.
\item {\it Generalizations to Imperfect Operations ---} Since the final
  criterion in a QBC protocol involves probabilities only, every step
  and requirement can also be relaxed to a probabilistic, rather than a
  perfect deterministic, one. For example, the verifying measurement
  by Babe need not succeed with probability 1. However, it seems that
  the relaxation should go only so far as to the case where the
  probability is arbitrarily close to 1, as in other quantum and
  classical algorithms. Even though there is no proof to the contrary,
  there is no known case where this particular generalization would ever lead
  to a US QBC protocol.
\end{itemize}

\paragraph{Generality of Quantum Purification ---} Adam needs to entangle
  his possible actions in (\ref{eq:entstate}) or
  (\ref{eq:multistage_equiv}) in order to launch an EPR attack as
  described in the IP. It has not been shown why all possible
  classically random elements in a protocol allow quantum entanglement
  purification. Indeed, when quantum teleportation involving one
  actual measurement among different possible spaces is utilized
  during commitment, as in the case of protocol QBC5 \cite{yuee},
  there is {\em no} quantum purification. In a different way, this situation
  of no purification, or no unique purification as
  in QBC1, may also obtain when a random part of a tensor
  product space from one party is to be returned to the other party,
  which occurs in many of our US protocols.\\

\paragraph{Different Commitment Possibilities ---} Under this category
  one may consider almost all the restrictions of the IP
  formulation that can be removed. A particularly important example is the
  possible use of multiple evidence state spaces. We have two separate secure protocols,
  QBC2 and QBC4, that exploit this possibility in different
  ways. Another example is the use of quantum teleportation in our QBC5.\\

\paragraph{Nonuniqueness ---}\label{par:nonunique} There are various places in the
IP where uniqueness of choice is implicitly assumed; otherwise the
question would arise as to why a cheating transformation $U^A$ can be
found which is successful for every possible choice. For example, the cheating probability
$\bar{P}^A_c$ depends on Babe's verifying measurement. For an
arbitrary protocol, the IP formulation does not, and
in fact cannot, specify what the possible verifying measurements could be. There is
{\em no} proof given that there cannot be more than one verifying
measurement for which different cheating transformations are
needed. When such a situation occurs, Adam may not know which one to
use for a successful cheating. However, this gap can be closed when the
verifying measurements are perfect, i.e., the bit is verified with
probability 1 from the measurement.

A more serious situation occurs in the case of purification
(\ref{eq:multistage_equiv}), when there is more than one way to purify a given
classical random number. For example, the usual multi-stage
formulation of QBC and coin tossing, exemplified in
(\ref{eq:multistage})-(\ref{eq:multistage_babe}), carries the implicit
assumption (and {\it restriction}) that one fixed space is passed
between the parties in the rounds. But there is great utility in splitting a classical or quantum correlated tensor product
space for obtaining security. If a random sample of $m$ out of $n$
qubits are to be sent from the first party to the second after state
modulation on the $m$ qubits, the first party can pick any $m$ of the
$n$ qubits and entangle/purify the result with unitary permutation
operators among all the $n$ qubits. Additional qubits apart from the
$n$ given ones can also be employed with proper permutation. As a
result, the $m$ qubits that are sent back can be any $m$-subset of the
original $n$ qubits plus other auxiliary qubits. However, the
resulting $\ket{\Phi_\sb}$ of (\ref{eq:multistage_equiv}) would be
different for different choices because the qubits have been
individuated by their positions, and there is no single overall
purification. Thus, there arises again the question of existence of a
uniformly successful cheating $U^A$ if concealing obtains in each of
the purifications. If the $m$ qubits cannot be entangled, or if their
entanglement cannot be maintained during the protocol, the possibility
of uniform cheating $U^A$ becomes the issue of the existence of
irreducible residual classical
randomness in the protocol that is answered negatively in our
following QBC1.

\section{Six new types of protocols}\label{sec:types}

In this section we describe six different types of protocols, together
with specific examples for five of them, that are not covered by the IP
for reasons expounded in the preceding section. For four protocols,
namely QBC1, QBC2, QBC4, and QBC5, full unconditional security proofs are
available. The situation for Type 3 and Type 6 protocols is not
certain. Together they should make clear the many possibilities that are open
for developing US QBC protocols. All our protocols assume for
simplicity that Adam
opens perfectly, i.e., with probability one, for $\sb=0$, as in the
IP.

\subsection{Type 1 protocols --- residual classical randomness}\label{ssec:type1}

Type 1 protocols are defined to be those in which there is inherent
classical randomness that cannot be quantum-mechanically purified and
maintained. This classical randomness is distinguished from that of a
Type 3 protocol that arises from $\{\lambda_k\}$ and $\{\ket{f_k}\}$
randomness in (\ref{eq:purif}). Our three-stage protocol QBC1 provides such an example and can be
motivated as follows. It is possible to create protocols that are
clearly binding; the question becomes how to make them concealing. The
main difficulty in this connection is Babe's entanglement (cheating)
over the random choices. This, it turns out, can be prevented during
cheating detection by Adam. Thus, the overall protocol becomes both
concealing and binding.

Consider a protocol in which Adam sends $n_0$ qubits to Babe, each
randomly drawn from a set of BB84 states $S = \{
\ket{1},\ket{2},\ket{3},\ket{4} \}$, $\braket{1}{3} =
\braket{2}{4}=0$, $\braket{1}{2} = \braket{3}{4}=1/\sqrt{2}$. Babe randomly picks one
and sends it back to Adam, who modulates it by $U_0 = I$ or $U_1 =
R_\pi$, the rotation by $\pi$ on the great circle containing $S$, and
commits it as evidence. He opens by telling Babe the state of each
qubit, which she verifies, telling $\sb=1$ if the one she sent was
moved by $R_\pi$. In a more complete protocol, Babe would check that Adam
indeed sends her states from $S$, and Adam would check that Babe is
sending back one of the qubits from him. This can be carried out
either in a classical game-theoretic formulation or through an ensemble approach
described in Appendices A and B of Ref.~\cite{yuee}. In any event, according to
IP and all coin-tossing formulations, both parties are assumed honest
except that they can (and should) entangle all possibilities.

The protocol is $\epsilon$-binding regardless of whether Babe entangles for the
following reason: Adam has to know exactly which qubit Babe sent back in
order to cheat successfully, regardless of whether or not he uses his
initial entanglement, which may involve permutations among the $n_0$
qubits. As just discussed in \ref{sec:ipprobs}d, IP does not apply
because he does not know which qubit Babe sends back. He needs to turn by $R_\pi$ just the one
qubit via $U^A$ that depends on exactly which qubit it is, i.e., on
the exact way Babe chose to purify and/or the exact $\ket{f_k}$ she
will measure. But if Adam knows which
qubit it is, he could just cheat by declaring a state appropriately
different from the original one, e.g., if he sent $\ket{3}$ he could
declare he sent $\ket{1}$ instead. Without knowing which qubit it is,
let $m (\le n)$ qubits be turned by him, each by an
amount that would be accepted as 1 with probability $p<1$ by Babe upon
her verification, while the other $n_0-m$ become a permutation of the original. (It can
be shown that with his full entanglement, the best he can do is to
turn a small fraction and re-permute the others, but this result is
not needed for the present argument.) Thus, his probability of
successful cheating is $(m/n_0)p(1-p)^{m-1}$, the maximum of which over
$m$ can be made arbitrarily small for large $n_0$.

If Babe does not entangle, this protocol is clearly perfectly
concealing. Since she may be able to cheat with permutation
entanglement, Adam can defeat that as follows. He sends her originally $n \gg n_0$
qubits from which Babe returns $n-n_0+1$ qubits. Adam randomly asks Babe
to reveal $n-n_0$ of the returned qubits and check that they are indeed
in states sent by him. The only entanglement Babe can employ is
permutation among the qubits as she could not respond to Adam
perfectly with additional entanglement. She may entangle a minimum of two
qubits at a time between one in the set of $n-n_0+1$ elements she sends back to
Adam and one in the set of $n_0-1$ elements she keeps, in order to maximize the
probability that the last one
retained by Adam remains entangled to at least one qubit in her
possession for her entanglement cheating. The probability of having
the entanglement surviving on the one remaining in Adam's possession
is $(n_0 -1)(n_0 - n +1)^{-1}$, which can be made arbitrarily
small. Hence, $\bar{P}^B_c < \frac{1}{2}+\epsilon$ for any $\epsilon$
and fixed $n_0$ with large $n$. This use of
$n \gg n_0$ qubits gives Adam new possibility of entanglement, which
he could not use under the protocol condition that he has to open the
bit on the one remaining qubit whose name Babe knows and would verify
upon. Thus, we have
proved that the following protocol is $\epsilon$-concealing and
$\epsilon$-binding.

\begin{center}
\vskip 0.12in
\framebox{
\begin{minipage}{0.9\columnwidth}
\vskip 0.1in
\underline{PROTOCOL {\bf QBC1}}

{\small \begin{enumerate}
\item[(i)] Adam sends Babe $n$ qubits, each drawn at random from the
  set $S$ of four BB84 states and named by its temporal position.
\item[(ii)]Babe randomly selects $n-n_0+1$ of them and sends them in a
  random order back to Adam, who asks Babe to reveal the names for
  $n-n_0$ of them. After verifying them, he modulates the remaining
  qubit for $U_0 = I$, $U_1 = R_\pi$.
\item[(iii)] Adam opens by declaring $\sb$ and the states of all the
  $n_0$ remaining qubits; Babe checks by corresponding projective measurements.
\end{enumerate}
\vskip 0.1in
}
\end{minipage}
}
\end{center}
\vskip 0.15in

To recapitulate the logic of its success, this protocol allows many
different purifications by Babe with different results
$\ket{\Phi_\sb}$ and which may not be concealing, so Adam cannot cheat anyway. However, by checking an ensemble Adam can force Babe
to measure and destroy her entanglement cheating possibility. The resulting protocol
becomes a classically randomized one, in which Adam of course still cannot cheat.

\subsection{Type 2 protocols --- bit-value dependent evidence state
  space}\label{ssec:type2}

As developed in Ref.~\cite{yued}, it is possible to have secure
protocols for which the evidence state space $\cH^B$ in $\cH^A
\otimes \cH^B$, which is in Babe's possession at the end of
commitment, depends on the bit $\sb$ and becomes $\cH^B_\sb$ as it appears to Adam, but is
indistinguishable for the two bit values to Babe. Type 2 protocols are
defined to be those in which Babe sends Adam $\cH^B_0$ and $\cH^B_1$
which she does not need to entangle to her kept spaces although she
may choose to. Adam returns $\cH^B_\sb$ to commit
$\sb$ while keeping the other space. It is distinguished from Type 4 protocols that employ
split entangled pairs to individuate $\cH^B_\sb$, and is easier to
implement practically. 

Protocol QBC2 goes as follows. Let Babe send Adam two sets of named states $S_0 = \{
\phi_{01},\ldots,\phi_{0n}\}$, $S_1 = \{\phi_{11},\ldots,\phi_{1n}\}$,
each $\phi_{\sb i}$ drawn randomly from the set $S$ of four BB84
states on a qubit. Adam does not know and cannot determine perfectly
what each state $\phi_{\sb i}$ is. To commit $\sb$, he sends back
randomly one of $\phi_{\sb i}$, revealing $\sb$ and $i$ at opening. In
order to cheat, Babe has to distinguish the two sets $S_0$ and $S_1$
and then measure on the committed state. It
is readily checked that this protocol is $\epsilon$-concealing for
sufficiently large $n$, such that the four states in $S$ appear in
nearly equal fractions among $S_0$ and $S_1$, even if Babe entangles
the states. One may impose the condition that each state in $S$
appears equally often in $S_0$ and $S_1$, which yields perfect
concealing if Babe does not entangle, but again only
$\epsilon$-concealing if she does. Such a condition may be obtained,
e.g., by sending in $n$ sets of four randomly permuted states instead,
with Adam picking one from a set. On the other hand, Adam cannot
cheat perfectly given that Babe does {\em not} entangle since that
does not really help her (or that she does but with $\{ \ket{f_k}\}$
of (\ref{eq:purif}) unknown to Adam; in the latter case it shares the
feature of QBC4 \cite{yued}). For QBC2 we just ake the case Babe does
not entangle. If Adam entangles the $\phi_{0i}$ by way of
permutations and commits a qubit $\cH_{0\bar{i}}$ or his own qubit $\cH_0$, he could {\em not}
change by local transformation the state in $\cH_{0\bar{i}}$ (or
$\cH_0$) to any of those in $\cH_{1i}$. That is, the state in
$\cH_{0\bar{i}}$ (or $\cH_0$)
is entangled with the states in $\cH_{0i}$ wherever these states may
go from local transformations. This is because of the invariance of
the position of a qubit within $S_0$ and $S_1$ under entanglement that
assures perfect $\sb=0$ opening. The IP does not apply because $U^A$
cannot be determined without knowing the state of the committed
$\phi_{\sb i}$. See the following Section~\ref{sec:concl} for further
discussion. Let $p_A < 1$ be Adam's optimal
cheating probability. As usual, this protocol QBC2.p can be
extended to an $\epsilon$-binding one, QBC2, in an $N$-sequence, making
$\bar{P}^A_c = p^N_A$ arbitrarily small.

\begin{center}
\vskip 0.12in
\framebox{
\begin{minipage}{0.9\columnwidth}
\vskip 0.1in
\underline{PROTOCOL {\bf QBC2}}

{\small \begin{enumerate}
\item[(i)] Babe sends Adam two $m$-qubit sets of states, $S_0$ and
  $S_1$, each state randomly drawn from a fixed set $S$ of four BB84
  states. To commit $\sb$, Adam randomly picks $N$ states, $N \ll m$,
  from $S_\sb$ and sends them to Babe.
\item[(ii)] Adam opens by revealing $\sb$ and all the states
  he committed; Babe verifies by projective measurements.
\end{enumerate}
\vskip 0.1in
}
\end{minipage}
}
\end{center}
\vskip 0.15in

\subsection{Type 3 protocols --- anonymous states}
\label{ssec:type3}

Type 3 protocols are defined to be those where concealing is obtained
for each of Babe's possible choices of $\{\lambda_k\}$ and/or
$\{\ket{f_k}\}$ in (\ref{eq:purif}) at any stage of the protocol. Each
choice thereby results in an {\em anonymous state} on $\cH^B_\alpha \otimes
\cH^B_\beta$ as it is unknown to Adam. To explain how such a situation may arise in view of the
purification (\ref{eq:purif}), observe that the unknown $\ket{\psi_k}$
without purification is merely replaced by the unknown $\{\ket{f_k}\}$
in (\ref{eq:purif}) even for known $\{\lambda_k\}$. How would the
other party, say Adam, know $\{\ket{f_k}\}$? Babe can use any
orthonormal $\{\ket{f_k}\}$ without affecting the protocol security,
assuming that the protocol is perfectly concealing for any orthonormal
$\{\ket{f_k}\}$, as it usually is. The Secrecy Principle \cite{yuec} mentioned above ensures that
Adam cannot demand to know exactly what $\{\ket{f_k}\}$ Babe uses in
any instance of the protocol execution. As a matter of fact, in
reality (\ref{eq:purif}) may just be an abstract representation such that
even Babe does not and cannot know $\{\ket{f_k}\}$, as for example
when Babe generates the $\{\ket{\psi_k}\}$ in a classically random
fashion. It is clear that IP would not go through unless Adam's
cheating transformation $U^A$ is independent of $\{ \ket{f_k}\}$. This
issue has not been examined in the literature, but a proof that such
independence is obtained was given in Ref.~\cite{yue4} for protocols
that do not involve what we call the switching of evidence state space
that produces the $\{\ket{f_k}\}$ dependence on $U^A$, or Babe's
checking over the entire entangled $U_{\sb i}\ket{\Phi}$ upon
verification that makes Theorem 3 of Ref.~\cite{yue4}
inapplicable. Moving the boundary to further entanglement on $\cH^C$
does not work because Adam cannot operate on $\cH^B_\beta$. Thus, the proof breaks down in general, and
the above scenario of $\{\ket{f_k}\}$ dependence of $U^A$ with
corresponding $U_{\sb i}\ket{\Psi}$ verification is carried out for the development of a secure
protocol QBC4.

In the early anonymous-state protocols \cite{yue3,yue4,yueb}, the use
of $\sb$-dependent evidence state spaces has not been discovered and
it was thought that $U^A$ is always independent of $\{\ket{f_k}\}$ in
(\ref{eq:purif}), even though a proof is only available for a special
class of protocols describe by $U_{\sb i}\ket{\Psi}$ or Eq.~(26) of
Ref.~\cite{yue4}. Furtheremore, the use of split entangled pair
verification on $U_{\sb i}\ket{\Psi}$ has also not been discovered,
and a theorem was proved in Ref.~\cite{yue4} that $U^A$ is independent
of $\{\lambda_k\}$ in the perfect concealing case. The question then
becomes whether the $\{\lambda_k\}$ freedom alone would yield a secure
protocol in the $\epsilon$-concealing case.

It is sometimes argued that such freedom cannot be
automatized and thus cannot lead to a definite QBC
protocol. However, since each party can clearly keep its own secret
mechanism of choosing the $\{ \lambda_k \}$, similar to other cases of
a kept secret in cryptography, this kind of protocols
are perfectly well-defined QBC protocols. In this connection, one
should avoid the confusion between a probability distribution $\{
\lambda_k \}$ and a definitive $\ket{\Psi}$ of (\ref{eq:purif}), and
between a random and an unknown parameter. See \cite{yuea}-\cite{yuec}
for further elaborations on these points. See also Ref.~\cite{ari} for
a general classification of many anonymous states protocols that,
however, does not include our four protocols of this paper.

Even if a theorem is proved with the purification (\ref{eq:purif}) for
any $\{\lambda_k\}$ and a fixed $\{\ket{f_k}\}$ that says $\epsilon$-concealing for all $\{\lambda_k\}$ yields a good cheating
$U^A$ independently of $\{\lambda_k\}$, it does not apply to our QBC2
where no $\{\ket{f_k}\}$ is used (or equivalently $U^A$ needs to succeed
for all $\{\ket{f_k}\}$) in Adam's cheating. This is because with
known $\{\ket{f_k}\}$, Adam does not need to really switch the state
from $S_0$ to $S_1$. Thus, our QBC2 protocol {\em is} a Type 3 protocol also
if one observes that Babe can entangle the $\cH^B_\alpha$ states
$\{\ket{\psi_k}\}$ with any $\cH^B_\beta$ state $\{\ket{f_k}\}$
without affecting $\epsilon$-concealing, but the cheating $U^A$
depends on $\{\ket{f_k}\}$.

\subsection{Type 4 protocols --- split entangled pair to individuate
  evidence space}\label{ssec:type4}

The use of bit-value dependent evidence state space leads to
$\epsilon$-concealing and $\epsilon$-binding protocol QBC2 described above. It is possible to obtain perfectly concealing
protocols when the evidence space is entangled by Babe,
which is indistinguishable to Babe when presented to her by Adam as committed evidence. This is described in
Ref.~\cite{yued} for protocol QBC4. These two protocols, although
relying on the same basic idea, illustrate in different ways the
diverse manifestation of $\sb$-dependent $\cH^B_\sb$ and classical
randomness in protocol design.

\subsection{Type 5 protocols --- utilizing quantum teleportation}\label{ssec:type5}

Type 5 protocols are defined to be those where quantum teleportation
is utilized during commitment. An example is a two-stage protocol in which Babe randomly
sends an entangled pair to Adam, who uses it to teleport a single
possible state $\ket{\sb}$ for each $\sb$ to one qubit and sends
it to Babe at opening.  The Bell measurement result is committed as
evidence. If Babe does not or cannot entangle, it is readily seen that
such a protocol is perfectly concealing, while Adam cannot cheat
perfectly. If Babe entangles her choice, the protocol can be made
$\epsilon$-concealing if Babe first sends in many qubit pairs. The resulting protocol QBC5p and its US extension QBC5 are
fully described in Ref.~\cite{yuee}. the main reason for the failure
of IP in this case is that Adam's measurement cannot be postponed to
after commitment while the possible measurement results cannot be
entangled as the actual reading is the committed evidence.

\subsection{Type 6 protocols --- necessary condition on concealing or
  binding}\label{ssec:type6}

Once the freedom of operation is opened up, one may no longer assume
that anything has to be known to both parties. For example, even in
the original IP formulation, one may allow Adam to use different
possible $\{ p_{\sb i} \}$ in (\ref{eq:entstate}). Thus, Babe has to
decide on $\sb$ by some strategy different from the one that assumes
$\{ p_{\sb i} \}$ are known. The issue of unknown versus random
parameters enters again, which greatly complicates the situation. It
is not clear how one may formulate necessary conditions for
concealing, or binding, that are needed to yield an impossibility
proof that says if the necessary concealing (or binding) condition is
satisfied, then the protocol cannot be binding (or
concealing). However, it is also not clear how to formulate a protocol
this way that can be proved secure. We just reserve the name ``Type 6
protocols'' for this approach without an example.

\section{Concluding remarks}\label{sec:concl}

We have tried to explain and to dissect the various ways a QBC
protocol can be designed, and to show the many possibilities that the
``impossibility proof'' formulation misses. Specifically, we list four
categories of gaps and six types of new protocol formulations that
exploit such gaps. In four of these types, there are protocols QBC1,
QBC2, QBC4, and QBC5 that can be proved unconditionally secure. This
also {\it solves} the quantum coin-tossing problem, in which much work has
been done assuming secure quantum bit commitment is
impossible. Also, one of our protocols, QBC2, can be readily
implemented with coherent states and is close to being practical even
with the limited quantum memory and quantum communication capabilities we
have at present. The challenge remains to find fully practical, secure QBC
schemes including system imperfections, as well as efficient ways to utilize them for various
cryptographic objectives.

\appendix

\section{Previous Protocols}

I discuss briefly here the security status of the protocols I previously
proposed and claimed unconditionally secure.

Protocol ``QBC2'' in Ref.~\cite{yue2}, the only one claimed to be US
there, is not proved perfectly concealing because entanglement
cheating by Babe is not accounted for. It was assumed to be the same
as a classically random protocol. While a proof is not available, it
appears that the protocol cannot be made perfectly concealing or
$\epsilon$-concealing without allowing Adam to cheat, even with the
modification described in Section~5 of Ref.~\cite{yuec}.

Ref.~\cite{yue3} is a preliminary version of Ref.~\cite{yue4}, in which
Type 3 protocols are introduced. However, the binding argument from
no-cloning there is not valid. Thus, if Adam knows the (purified)
anonymous state, the protocol is insecure due to the fact that all
perfectly verifying measurements under the IP formulation lead to
insecure protocols against a single cheating $U^A$, although that is a fact not proved in
the IP itself. If Adam does not know the anonymous state, the
situation is not yet completely resolved. However, it appears that
Adam can probably cheat for operator-theoretic reasons on tensor
product spaces, in this case as well as in QBC2.1p of v2 of this paper.

Protocols ``QBC1'' in Ref.~\cite{yuea} and ``QBC4'' in Ref.~\cite{yueb},
as well as the preliminary version of Ref.~\cite{yue2}, involve attemps to force Adam to measure on $\cH^A$, thus effectively
``destroyng'' his entanglement. They fail as indicated in Section
\ref{sec:impproof}. However, such an attempt can succeed when applied
to destroy Babe's entanglement, in the sense discussed in this paper, which leads to our present QBC1 that was first briefly
discussed in Ref.~\cite{yuec}. The protocol ``QBC2'' in
Ref.~\cite{yuea} and v1 of this paper, which is a ``Type 2'' protocol
of Ref.~\cite{yuec}, protocl ``QBC4'' in v1 of
Ref.~\cite{yuee}, and protocol ``QBC5'' in v1 of Ref.~\cite{yued} are all
insecure because they can be brought into the form of a single
$\ket{\Phi_\sb}$ so that IP and the results of Section~III in
Ref.~\cite{yue4} apply.


\begin{acknowledgments}

I would like to thank G.M.~D'Ariano, W.Y.~Hwang, H.K.~Lo, R.~Nair, and
M.~Raginsky for useful discussions. This work was supported by the Defense Advanced Research Projects
Agency and by the U.S. Army Research Office.

\end{acknowledgments}

\end{document}